\documentclass[twocolumn]{aastex62}
\usepackage{graphicx}
\usepackage{natbib}
\usepackage{xcolor}
\shorttitle{Nebular \ha~Limits}
\shortauthors{Sand et al.}

\newcommand{\ha}{H$\alpha$}

\begin{document} 

\title{Nebular H$\alpha$ Limits for Fast Declining Type Ia Supernovae}

\author[0000-0003-4102-380X]{D. J. Sand}
\affiliation{Steward Observatory, University of Arizona, 933 North Cherry Avenue, Tucson, AZ 85721-0065, USA}

\author[0000-0002-1546-9763]{R. C. Amaro}
\affiliation{Steward Observatory, University of Arizona, 933 North Cherry Avenue, Tucson, AZ 85721-0065, USA}

\author{M. Moe}
\affiliation{Steward Observatory, University of Arizona, 933 North Cherry Avenue, Tucson, AZ 85721-0065, USA}

\author[0000-0002-9154-3136]{M. L. Graham}
\affiliation{Department of Astronomy, University of Washington, Box 351580, U.W., Seattle, WA 98195-1580, USA}

\author[0000-0003-0123-0062]{J.E. Andrews}
\affiliation{Steward Observatory, University of Arizona, 933 North Cherry Avenue, Tucson, AZ 85721-0065, USA}

\author{J. Burke}
\affiliation{Department of Physics, University of California, Santa Barbara, CA 93106-9530, USA}
\affiliation{Las Cumbres Observatory, 6740 Cortona Dr, Suite 102, Goleta, CA 93117-5575, USA}

\author{R. Cartier}
\affiliation{Millennium Institute of Astrophysics, Casilla 36-D, Santiago, Chile}
\affiliation{Departamento de Astronomía, Universidad de Chile, Casilla 36-D, Santiago, Chile}

\author{Y. Eweis}
\affiliation{Department of Physics and Astronomy, Rutgers, the State University of New Jersey, 136 Frelinghuysen Road, Piscataway, NJ 08854, USA}

\author[0000-0002-1296-6887]{L. Galbany}
\affiliation{PITT PACC, Department of Physics and Astronomy, University of Pittsburgh, Pittsburgh, PA 15260, USA}

\author[0000-0002-1125-9187]{D. Hiramatsu}
\affiliation{Department of Physics, University of California, Santa Barbara, CA 93106-9530, USA}
\affiliation{Las Cumbres Observatory, 6740 Cortona Dr, Suite 102, Goleta, CA 93117-5575, USA}

\author{D. A. Howell}
\affiliation{Department of Physics, University of California, Santa Barbara, CA 93106-9530, USA}
\affiliation{Las Cumbres Observatory, 6740 Cortona Dr, Suite 102, Goleta, CA
 93117-5575, USA}

\author[0000-0001-8738-6011]{S. W. Jha}
\affiliation{Department of Physics and Astronomy, Rutgers, the State University of New Jersey, 136 Frelinghuysen Road, Piscataway, NJ 08854, USA}
\affiliation{Center for Computational Astrophysics, Flatiron Institute, 162 5th Avenue, New York, NY 10010, USA}

\author{M. Lundquist}
\affiliation{Steward Observatory, University of Arizona, 933 North Cherry Avenue, Tucson, AZ 85721-0065, USA}

\author[0000-0001-6685-0479]{T. Matheson}
\affiliation{National Optical Astronomy Observatory, 950 North Cherry Avenue, Tucson, AZ 85719, USA}

\author{C. McCully}
\affiliation{Department of Physics, University of California, Santa Barbara, CA 93106-9530, USA}
\affiliation{Las Cumbres Observatory, 6740 Cortona Dr, Suite 102, Goleta, CA 93117-5575, USA}

\author{P. Milne}
\affiliation{Steward Observatory, University of Arizona, 933 North Cherry Avenue, Tucson, AZ 85721-0065, USA}

\author[0000-0001-5510-2424]{Nathan Smith}
\affiliation{Steward Observatory, University of Arizona, 933 North Cherry Avenue, Tucson, AZ 85721-0065, USA}

\author{S. Valenti}
\affiliation{Department of Physics, University of California, 1 Shields Avenue, Davis, CA 95616-5270, USA}

\author{S. Wyatt}
\affil{Department of Astronomy/Steward Observatory, 933 North Cherry Avenue, Rm. N204, Tucson, AZ 85721-0065, USA}


\begin{abstract}
One clear observational prediction of the single degenerate progenitor scenario as the origin of type Ia supernovae (SNe) is the presence of relatively narrow ($\approx$1000 km s$^{-1}$) H$\alpha$ emission at nebular phases, although this feature is rarely seen.  We present a compilation of nebular phase H$\alpha$ limits for SN Ia in the literature and demonstrate that this heterogenous sample has been biased towards SN Ia with relatively high luminosities and slow decline rates, as parameterized by $\Delta$m$_{15}(B)$, the difference in $B$-band magnitude between maximum light and fifteen days afterward.  Motivated by the need to explore the full parameter space of SN~Ia and their subtypes, we present two new and six previously published nebular spectra of SN Ia with $\Delta$m$_{15}(B)$$~>~$1.3 mag (including members of the transitional and SN1991bg-like subclasses) and measure nondetection limits of $L_{H\alpha}$$~<~$0.85--9.9$\times$10$^{36}$ ergs s$^{-1}$, which we confirmed by implanting simulated H$\alpha$ emission into our data.  Based on the lastest models of swept-up material stripped from a nondegenerate companion star, these  $L_{H\alpha}$ values correspond to hydrogen mass limits of $M_H$$~\lesssim~$1-3$\times$10$^{-4}$ $M_{\odot}$, roughly three orders of magnitude below that expected for the systems modeled, although we note that no simulations of H$\alpha$ nebular emission in such weak explosions have yet been performed.  Despite the recent detection of strong H$\alpha$ in ASASSN-18tb (SN~2018fhw; $\Delta$m$_{15}(B)$~=~2.0 mag), we see no evidence that fast declining systems are more likely to have late time H$\alpha$ emission, although a larger sample is needed to confirm this result.
\end{abstract}

\keywords{supernovae: general}

\section{Introduction}
\label{sec:intro}

Despite their intense study and use as standardizable candles to measure the expansion history of the universe, the exact nature of the progenitor system(s) of type Ia supernovae (SNe Ia) is still not known.  Two main progenitor channels are considered -- the single degenerate and double degenerate scenarios.  In the single degenerate (SD) scenario, a single carbon-oxygen white dwarf gains material from a nondegenerate companion star \citep{Whelan73}, while in the double degenerate (DD) scenario there is a second degenerate star in the system \citep{Iben84,Webbink84}.  The details of how the thermonuclear explosion is triggered in each of these configurations is being actively studied. 

There are several observational signatures that can signal the presence of a nondegenerate companion star and the single degenerate scenario \citep[for a review, see e.g.,][]{Maoz14}, but here we focus on the presence of narrow  hydrogen lines in nebular phase SN Ia spectra.  In the single degenerate scenario, the SN ejecta can collide with the companion star and strip its surface of material.  Once the SN emission becomes optically thin at late times, and the observer can see into the central regions of the explosion, this swept up material manifests itself as a relatively strong and narrow emission line.  Initially this emission was anticipated in \citet{Marietta00}, and has since been the subject of several theoretical and numerical studies \citep[e.g.,][]{Mattila05,Pan12,Liu13,Lundqvist13,Boty18}.  The models generically show narrow ($\sim$1000 km s$^{-1}$) emission from H$\alpha$ in particular (helium and other lines are also plausible, depending on the companion star, although we focus on hydrogen in the current work) but there is considerable diversity in the predicted strength of the line from study to study, which depends on the physics included in each simulation, the strength of the explosion, and  the details of the companion separation and type.  This parameter space has not yet been explored in the models.

Definitive late-time narrow emission lines have not been detected in standard SN Ia, despite observations of 20 systems   \citep[see Table~\ref{tab:litsearch2}; ][ and see the recently submitted \citealt{Tucker19} for even more systems]{Mattila05,Leonard07,Shappee13,Lundqvist13,Lundqvist15,Maguire16,Graham17,Shappee18,Holmbo18,
Sand18_2017cbv,Tucker18,Dimitriadis19}.  We note that there is a class of SN Ia-like objects, the so-called SN~2002ic-like or SN Ia-CSM events \citep[e.g.,][]{Hamuy03,Silverman13}, that display narrow H$\alpha$ emission from early on in the SN's evolution \citep[but see][]{Dilday12,Graham19}.  This emission is thought  to originate from interaction with circumstellar material, although it is possible they are a signpost of the single degenerate scenario.  We largely leave aside these objects for the present study, although we note that they generally sit on the bright end of the SN Ia distribution, and have slowly declining light curves \citep[e.g.,][see also Figure 1 of \citealt{Taubenberger17}]{Leloudas15}.



Recently, \citet{Kollmeier19} presented a nebular spectrum (+139d after maximum light) of ASASSN-18tb (SN~2018fhw), a fast-declining ($\Delta$m$_{15}(B)$~=~2.0 mag) and subluminous SN Ia hosted by a dwarf elliptical galaxy.  ASASSN-18tb was discovered by the All Sky Automated Survey for Supernovae \citep[ASASSN;][]{Shappee14}, and is studied in further detail by \citet{Vallely19_18tb}. The spectrum showed clear, conspicuous $H\alpha$ emission with a luminosity of $L_{H\alpha}$~=~2.2$\times$10$^{38}$ ergs s$^{-1}$ and FWHM$\approx$1100 km~s$^{-1}$.  In the context of the latest radiative transfer models of the expected hydrogen emission signature in nebular spectra \citep{Boty18}, such an H$\alpha$ luminosity corresponds to $\sim$2$\times$10$^{-3}$ M$_{\odot}$ of swept up material, below expectations for the single degenerate model.  Nonetheless, uncertainties in the modeling and the fact that the models thus far have not studied subluminous SN Ia make this and other scenarios worth exploring.  Further data on ASASSN-18tb are also needed to determine if other observational signatures of the single degenerate model are evident.

Motivated by the H$\alpha$ emission and fast declining nature of ASASSN-18tb, in Section~\ref{sec:sample} we compile the literature sample of SN Ia with nebular H$\alpha$ limits as a function of their light curve decline rate, $\Delta$m$_{15}(B)$, which is known to correlate with the luminosity and color of SN Ia \citep{Phillips93}.   From this compiled dataset, we notice a lack of H$\alpha$ limits for SN Ia with a decline rate of $\Delta$m$_{15}(B)$$~\gtrsim~$1.3 mag.  We then gather a sample of both new and archival nebular SN Ia spectra to constrain the incidence of late time H$\alpha$ in such fast declining systems, partially to explore this new parameter space and to put the recent results for ASASSN-18tb in context.

\begin{figure}
	\centering
	\includegraphics[width=2.9in]{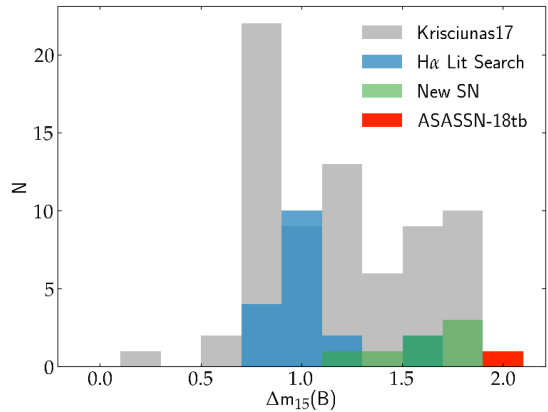}
	\caption{Histogram of $\Delta m_{15}(B)$ values for SNe Ia with nebular H$\alpha$ limits from the literature (blue; see also Table~\ref{tab:litsearch2}) in comparison to the Carnegie Supernova Project's sample \citep[grey; ][]{Krisciunas17}. The sample with H$\alpha$ limits is clearly biased towards SN Ia with lower $\Delta m_{15}(B)$ values (the brighter and slower declining events).  We also show ASASSN-18tb (red), which has a clear H$\alpha$ detection at $\Delta m_{15}(B)$~=~2.0 mag, an area of parameter space that has not yet been sampled.  The green histogram represents the new measurements presented in this work. 
	\label{fig:m15hist}}
\end{figure}

\section{Nebular H$\alpha$ Literature Search}\label{sec:sample}


We have collected all of the H$\alpha$ nebular spectroscopy limits for SN Ia reported in the literature \citep[][]{Mattila05,Leonard07,Shappee13,Lundqvist13,Lundqvist15,Maguire16,Graham17,Shappee18,Holmbo18,
Sand18_2017cbv,Tucker18,Dimitriadis19}, as well as the real detection of ASASSN-18tb \citep{Kollmeier19}, and listed them in Table~\ref{tab:litsearch2} under the `From Literature' portion of the table.  For each object we also list the most constraining H$\alpha$ flux and luminosity limits, along with the phase at which the data was taken, labeled in days since $B$-band maximum.    These results have been taken directly from the original papers and we make no attempt to homogenize them, and one should keep in mind that many studies use different methodologies for deriving their observational limits.  

At the same time, we have looked at the distribution in decline rates among these SN Ia, parameterized by $\Delta$m$_{15}(B)$, the difference in magnitude between a SN Ia at peak and 15 days after maximum in the $B$-band, which has long been known to correlate with the peak luminosity and color of SNe Ia \citep{Phillips93}.  The fastest declining SN Ia have the faintest absolute magnitudes, and vice versa.   

In Figure~\ref{fig:m15hist} we plot the histogram of  $\Delta$m$_{15}(B)$ values for the nebular H$\alpha$ sample in comparison to the Carnegie Supernova Project's (CSP) sample of SN Ia \citep{Krisciunas17}, which we assume is nearly representative of the population as a whole.  The CSP attempted to observe nearly every low redshift SN in the southern hemisphere over a 5-year period during their observing season \citep[see also][]{Hamuy06}, although these SNe were largely drawn from pointed galaxy SN searches, which was all that was available at the time.  We also plot ASASSN-18tb, the recently announced SN Ia with nebular H$\alpha$ emission.  There are several things to note.  First, ASASSN-18tb is on the very upper edge of the distribution of SN Ia in terms of $\Delta$m$_{15}(B)$.  Second, the population of SN Ia that have nebular H$\alpha$ limits is biased toward slower declining, relatively brighter events.  There is a lack of H$\alpha$ limits at $\Delta$m$_{15}(B)~\gtrsim~$1.3 mag, while the recent results for ASASSN-18tb indicate that is a promising region of parameter space to search for this signature of the SD scenario.  In the next section, we attempt to further  populate the $\Delta$m$_{15}(B)~\gtrsim~$1.3 mag region of parameter space by both presenting new SN~Ia nebular spectra and measuring limits for previously published spectra that do not yet have reported H$\alpha$ measurements.  

\begin{deluxetable*}{l c c c c c c}
\tabletypesize{\scriptsize}
\tablecaption{SN Ia Nebular H$\alpha$ limits and light curve parameters.  The top portion of the table is a compilation of literature values taken directly from their source, while the bottom presents new measurements of SN Ia with $\Delta m_{15}(B)$ $~>~$1.3 mag. \label{tab:litsearch2}}
\tablehead{\colhead{SN Name} & \colhead{$\Delta$m$_{15}(B)$} & \colhead{\ha~Flux Limit}  & \colhead{\ha~Luminosity Limit} & \colhead{Phase\tablenotemark{*}} & \colhead{$\Delta$m$_{15}(B)$ ref.}  & \colhead{\ha~ref.} \\ 
 & \colhead{(mag)} & \colhead{(10$^{-17}$ erg s$^{-1}$ cm$^{-2}$)} & \colhead{(erg s$^{-1}$)} & \colhead{(days)} & &  }
\startdata
\hline
\multicolumn{7}{c}{From Literature}\\
\hline
SN~2001el & 1.13 & ...\tablenotemark{a} & ...\tablenotemark{a} & +398 & \cite{Krisciunas03} & \cite{Mattila05}\\
SN~2005am & 1.73 & 0.172 & $2.7 \times 10^{35}$ & +381 & \cite{Hicken09} & \cite{Leonard07} \\ 
SN~2005cf & 1.06 & 0.223 & $2.7 \times 10^{35}$ & +384 & \cite{Hicken09} & \cite{Leonard07} \\ 
SN~2009ig & 0.89 & 0.33 & $4.7 \times 10^{35}$ & +405 & \cite{Foley12a} & \cite{Maguire16} \\ 
SN~2010gp & 1.19 & 0.18 & $2.4 \times 10^{36}$ & +277 & \cite{Brown17} & \cite{Maguire16} \\ 
SN~2011ek & 1.13\tablenotemark{b} & 0.18 & $6.2 \times 10^{34}$ & +421 & \cite{Sullivan12} & \cite{Maguire16} \\ 
SN~2011fe & 1.18 & 3.14 & $1.5 \times 10^{35}$ & +274 & \cite{Zhang16} & \cite{Shappee13}\tablenotemark{c} \\ 
SN~2011iv & 1.69 & 1.87 & $9.7 \times 10^{35}$ & +318 & \cite{Foley12b} & \cite{Maguire16} \\ 
SN~2012cg & 0.98 & 3.09 & $8.5 \times 10^{35}$ & +339 & \cite{Vinko18} & \cite{Maguire16}\tablenotemark{d} \\ 
SN~2012cu & 1.05\tablenotemark{b} & 0.56 & $1.3 \times 10^{36}$ & +344 & \cite{Amanullah15} & \cite{Maguire16} \\ 
SN~2012fr & 0.82 & 4.79 & $2.1 \times 10^{36}$ & +357 & \cite{Contreras18} & \cite{Maguire16} \\ 
SN~2012ht & 1.39 & 2.05 & $9.9 \times 10^{35}$ & +433 & \cite{Yamanaka14} & \cite{Maguire16} \\ 
SN~2013aa & 1.02 & 1.46 & $5.7 \times 10^{35}$ & +360 & \cite{Graham17} & \cite{Maguire16} \\ 
SN~2013cs & 1.11 & 1.18 & $2.7 \times 10^{36}$ & +303 & \cite{Childress16} & \cite{Maguire16} \\ 
SN~2013ct\tablenotemark{e} & ... & 5.66 & $9.9 \times 10^{35}$ & +229 & ... & \cite{Maguire16} \\ 
SN~2013gy & 1.23 & 45.0 & $1.7 \times 10^{38}$ & +235 & \cite{Holmbo18} & \cite{Holmbo18} \\ 
SN~2014J & 1.12 & 16.0 & $2.2 \times 10^{35}$ & +315 & \cite{Marion15} & \cite{Lundqvist15} \\ 
SN~2017cbv & 1.06 & 4.40 & $8.0\times10^{35}$ & +302 & \cite{Hosseinzadeh17} & \cite{Sand18_2017cbv} \\
SN~2018oh & 0.96 & 0.32 & $2.6\times10^{37}$ & +236 & \cite{Li19}& \cite{Dimitriadis19}\tablenotemark{f}  \\
ASASSN-18tb\tablenotemark{g} & 2.0 & 35.0 & $2.2\times10^{38}$ & +139 & \cite{Kollmeier19} & \cite{Kollmeier19} \\
\hline
\multicolumn{7}{c}{New Measurements}\\
\hline
SN~1999by  & 1.90 & 22.3 & $5.3\times10^{36}$ & +183  & \cite{Garnavich04} & This work\\
SN~2003hv & 1.61 & 2.0 & $8.5\times10^{35}$ & +320 & \cite{Leloudas09} & ''\\
SN~2003gs & 1.83 & 15.5 & $8.4\times10^{36}$ & +200 & \cite{Krisciunas09} & ''\\
SN~2004eo & 1.46 & 1.8 & $9.9\times10^{36}$ & +228 & \cite{Pastorello07} & ''\\
SN~2007gi & 1.40 & 4.2 & $3.0\times10^{36}$ & +225 & \cite{Zhang10} & ''\\
SN~2007on & 1.96 & 8.2 &$3.1\times10^{36}$ & +286 & \cite{Gall18} & '' \\
SN~2016brx\tablenotemark{h} & ... & 2.0 & $8.2\times10^{36}$ & +183 & ... & ''\\
SN~2017fzw & 1.79 & 11.3 & 8.8 $\times$10$^{36}$ & +233 & Galbany et al. in prep & ''\\
\enddata
\tablenotetext{*}{Phases are with respect to $B$-band maximum.}
\tablenotetext{a}{\citet{Mattila05} report an informal limit of $M_H$ $\approx$ 0.03 $M_{\odot}$ but no explicit flux/luminosity limit.}
\tablenotetext{b}{\label{stretch}Converted from stretch using the method from \cite{Altavilla04}}
\tablenotetext{c}{See also \cite{Lundqvist15}}
\tablenotetext{d}{See also \cite{Shappee18}}
\tablenotetext{e}{No light curve information is available for SN~2013ct}
\tablenotetext{f}{See also \citet{Tucker18}}
\tablenotetext{g}{ASASSN-18tb had a clear detection of H$\alpha$, unlike the other objects in this table.  The reported values in this row are the true measurements, not limits.}
\tablenotetext{h}{While SN~2016brx does not have a measured $\Delta$m$_{15}(B)$ we place it in our sample as it is a confirmed SN1991bg-like, fast declining SN Ia \citep{Dong18}.}
\end{deluxetable*}

\section{Nebular Spectroscopy of Fast Declining SN Ia \label{sec:data}}

In this section we present both new and archival nebular spectroscopy of SN Ia that have a $\Delta$m$_{15}(B)~>~$1.3 mag in order to fill in some of the unexplored parameter space seen in Figure~\ref{fig:m15hist}; we further discuss the broad range of SNIa and their subtypes that this sample represents below.  For clarity, we refer to nebular spectra that have been previously published, but have not been used for placing limits on H$\alpha$ as `archival' in this work, and they should not be confused with the H$\alpha$  literature results presented in the top half of Table~\ref{tab:litsearch2}. Both the new and archival data are from multiple programs with different science goals, and so the dataset is heterogenous in wavelength coverage, resolution, and signal-to-noise ratio.  

The archival dataset is a result of a literature search of prominent SN Ia that have been studied in detail, and have reduced spectra in public databases \citep{Silverman12,Yaron12,OpenSN}.   No attempt was made to search telescope archives for unpublished data, although such a search would likely be fruitful.  All spectra must be at least +130 days past maximum in the $B$ band in order to qualify as a nebular spectrum, and have a published $\Delta$m$_{15}(B)$ value.  We made one exception by including the SN~2016brx into our sample, which does not have a $\Delta$m$_{15}(B)$ measurement due to sparse light curve coverage, although it is a bona fide fast declining SN1991bg-like SN Ia that would make it into our sample if the light curve data existed; see \citet{Dong18} for details. If a previously published SN had multiple nebular spectra available, we only chose the highest signal to noise example to include in this study.  We did not include objects whose nebular spectra were evidently low in quality upon inspection, or had other relevant issues.  For instance, SN~1986G would be an excellent addition to our sample \citep[$\Delta$m$_{15}(B)$=1.62;][]{Phillips87}, but due to its proximity to an HII region many of its nebular spectra have  oversubtraction issues near H$\alpha$ \citep{Cristiani92}, and so we excluded it from our sample.
During our search, Table~8 in \citet{Hsiao15} and Table~3 in \citet{Vallely19} both proved useful for identifying SNe Ia that matched our search criteria described above.

The upper portion of Table~\ref{tab:newspec} shows the observation log for the six spectra identified in the literature in our archival data set, and the data themselves are shown in Figure~\ref{fig:spectra}.  We also list other parameters essential for inferring H$\alpha$ limits in this table, including the derived color excess and distance to each SN.  We emphasize again that this is not likely a complete set of published nebular SN Ia spectra with $\Delta$m$_{15}(B)~>~$1.3 mag without H$\alpha$ limits. A more comprehensive search may yet turn up more examples.

We also present new nebular-phase spectra for two SN Ia, which are detailed in the bottom portion of Table~\ref{tab:newspec}.  All of the spectra were reduced in a standard way, performing bias subtraction, flat fielding, cosmic ray rejection, local sky subtraction and extraction of one-dimensional spectra.  A sensitivity function was derived from standard star observations to flux calibrate the spectra, which are presented in Figure~\ref{fig:spectra}.

In order to account for nonphotometric conditions and slit losses, and to get meaningful $H\alpha$ narrow line limits, it is necessary to place our spectra on an absolute scale.  For both the new and archival data, we rescale the spectra
to match late time photometry at a given epoch via simple linear interpolation or mild extrapolation.  The source of the late time light curve is detailed in Table~\ref{tab:newspec}, as well as the magnitude that the spectrum was scaled to.  One exception is SN~2016brx, which has a very sparse late time light curve, but which was matched with a SN1991bg light curve template by \citet{Dong18}; we estimated a R=22.4 mag at the time of the nebular spectrum at +184d based on this comparison\footnote{On Nov. 4, 2016 (+199d), we also obtained 3$\times$15-min V-band exposures of SN2016brx with the 61-in Kuiper Telescope and Mont4K imager, and measured V = 23.0$\pm$0.2 mag.  Assuming that 91bg-like objects have a color of $V$$-$$R$$\approx$0 during this time period, and they decline at a rate of 0.025 mag per day \citep{Milne01}, than we would have expected $R$=22.6$\pm$0.2 mag at +184d, confirming our estimate of R $\approx$ 22.4 mag at the same epoch.}.
Another exception is SN~2007on, which does not have light curve data beyond +90 days after $B$-band maximum.  For this SN, we instead used the late time light curves of SN2007gi and SN2003gs (both in our sample), rescaling them to the distance of SN~2007on and taking the average result between the two, resulting in an inferred $R$~=~21.59 mag for SN~2007on at +286 days; the light curves were all well-matched at earlier epochs.

No previous data has been published for  SN~2017fzw, and so we measure it's light curve parameters directly (Tables 1 \& 2).  This SN Ia was discovered by the Distance Less Than 40 Mpc (DLT40) fast cadence nearby SN search \citep{Tartaglia18}, and its full dataset -- including further work on its nebular spectra --  will be studied elsewhere (Galbany et al., in preparation).  We simply provide H$\alpha$ detection limits here. 

Once the spectra have been scaled to the photometry, we then correct for any Milky Way and host extinction using the extinction model of \citet{Fitzpatrick99}.  The values for the color excess used for this correction are listed in Table~\ref{tab:newspec}.

The properties of our new SN Ia nebular sample can be gleaned from their decline rates.  They range from relatively normal SN Ia on the fast end of the decline rate distribution (e.g. SN~2007gi with $\Delta$m$_{15}(B)$~=~1.40 mag; \citealt{Zhang10}) to so-called transitional SN Ia (e.g. SN~2007on with $\Delta$m$_{15}(B)$~=~1.90 mag; \citealt{Gall18}), which display properties intermediate to that of normal SN Ia and SN 91bg-like events.  The SN~1999by belongs to the class of SN~91bg-like events, given its lack of secondary maximum, and conspicuous \ion{Ti}{2} in its optical spectrum \citep{Garnavich04}; as mentioned previously, SN~2016brx also belongs to this class \citep{Dong18}.  SN~2017fzw has secondary maxima in its $i$-band light curve, and so is not a true SN 91bg-like event \citep[see discussion in ][]{Gall18}; we postpone a more detailed discussion of SN~2017fzw's properties to future work.  While it is beyond the scope of the current work, it is possible that fast declining normal SN Ia, transitional SN Ia and SN-91bg-like SNe all have different progenitor systems, and so more detailed studies of the nebular H$\alpha$ statistics of each subclass may be warranted \citep[e.g. see discussions in][]{Hsiao15,Gall18,Dhawan17}. At the time of this writing, it is not clear if ASASSN-18tb fits in cleanly into any of these subtypes, although it seems to have properties of both the transitional and SN91bg-like SN Ia -- we discuss this system further in Section~\ref{sec:conc}.  Nonetheless, as seen in Figure~\ref{fig:m15hist}, it is clear that H$\alpha$ limits are needed for all subtypes at the fast decline end of the SN Ia distribution.




\section{Stripped Material Search via H$\alpha$}\label{sec:search}

\begin{figure*}
\begin{center}
\includegraphics[scale=0.62]{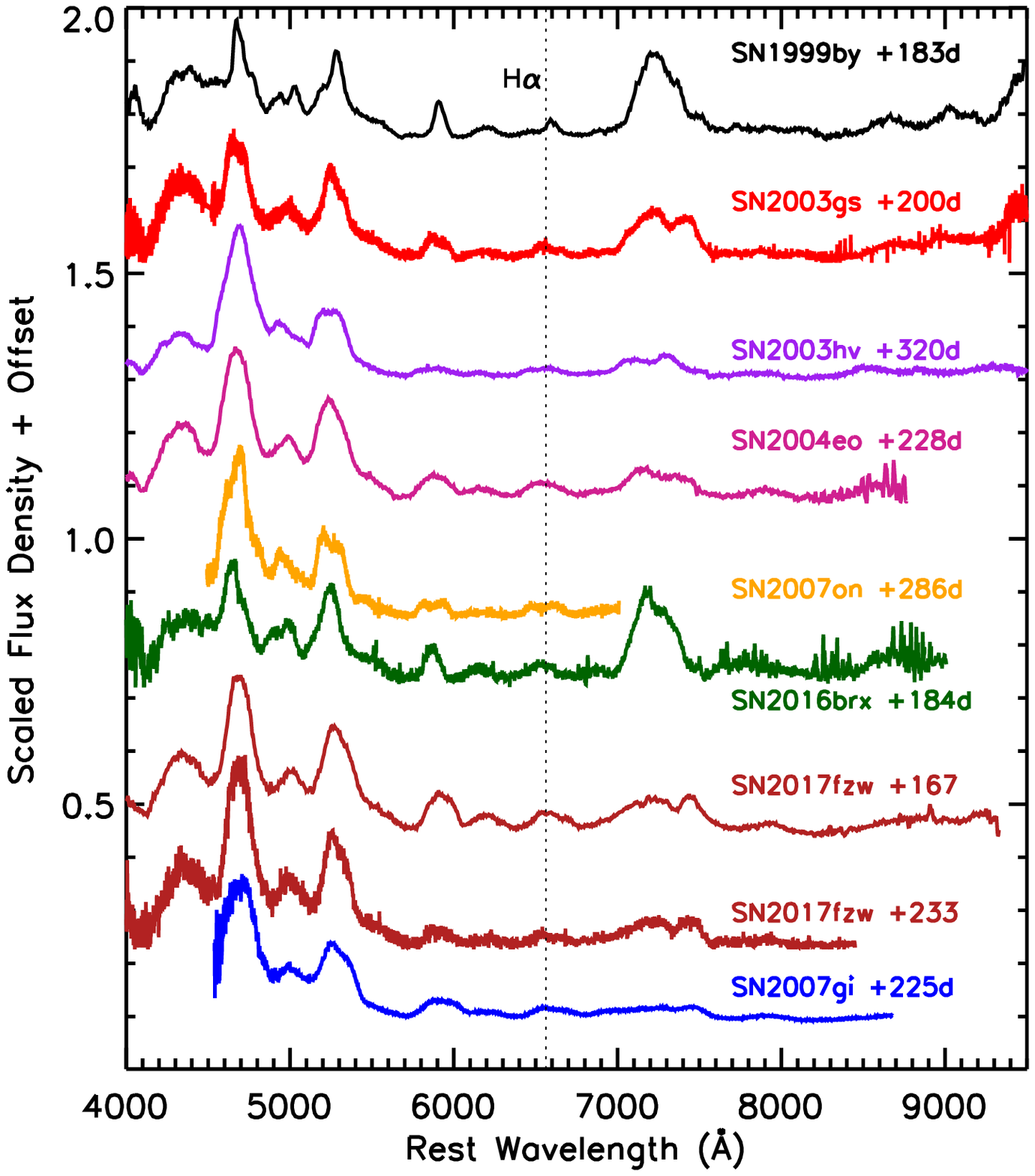}
\includegraphics[scale=0.62]{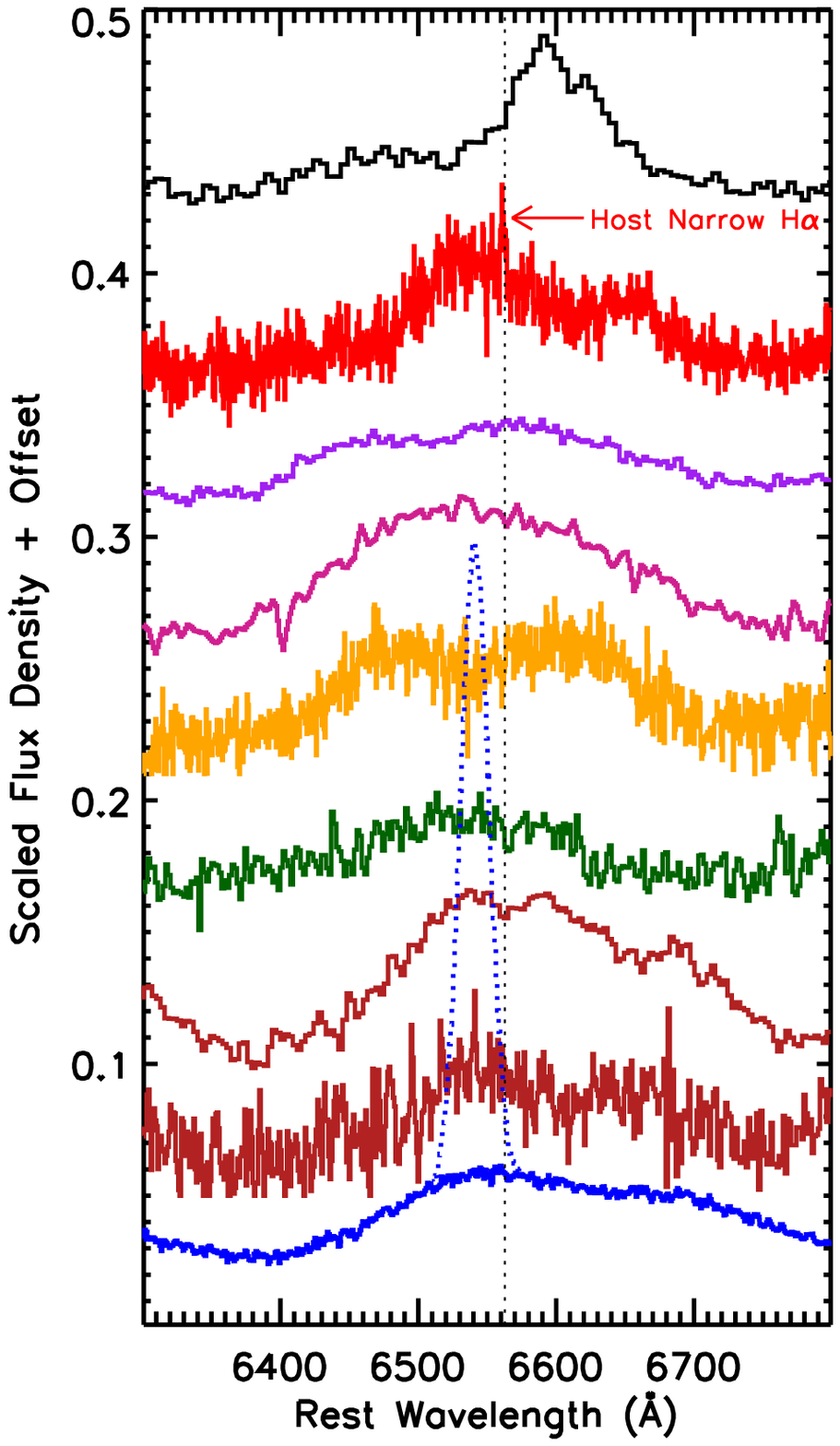}
\caption{Nebular phase spectra of SNe Ia in our $\Delta$m$_{15}(B)~>~$1.3 mag sample; the spectra have been rescaled for presentation purposes.   Spectra are labeled in the left panel, and have the same colors in both panels.  On the right is a zoom in of the region around H$\alpha$, where one would expect narrow (FWHM$\approx$1000 km s$^{-1}$) emission to be apparent in the single degenerate scenario, although we see no such signature.  For illustration, we have placed a simulated H$\alpha$ line with FWHM=1000 km s$^{-1}$, and offset by $-$1000 km s$^{-1}$, onto the spectrum of SN~2007gi.  This simulated feature has the same luminosity as the H$\alpha$ line seen in ASASSN-18tb -- we clearly see no such features in our sample. We do mark a very narrow H$\alpha$ feature in SN~2003gs, which has a width of $\approx$75 km~s$^{-1}$.  This feature is near the resolution of the Keck/ESI spectrum.  Upon inspection of the 2D spectrum, this H$\alpha$ emission clearly originates from the host galaxy rather than the SN; see Section~\ref{sec:search} for details.
\label{fig:spectra}}
\end{center}
\end{figure*}

As discussed, a clear observational signature of the single degenerate scenario is that stripped hydrogen (and possibly helium) rich material would be swept up and cause narrow emission lines ($\sim$1000 km s$^{-1}$) at late times.  Here we focus on the H$\alpha$ line, which is predicted to be strongest and is the subject of most modeling efforts.  From inspection of Figure~\ref{fig:spectra}, there are no strong, narrow H$\alpha$ lines apparent.  We note that during the nebular phase, an H$\alpha$ feature would sit atop the broad [\ion{Co}{3}] emission feature seen in the zoomed in portion of this figure.  The exception is SN~2003gs, which has an apparent H$\alpha$ emission feature which is narrower than expected from the single degenerate scenario, which we discuss further below.  We focus on setting limits on any trace amounts of H$\alpha$ in these spectra, assuming a FWHM~=~1000 km s$^{-1}$ and a potential offset from the rest wavelength of up to $\sim$1000 km s$^{-1}$, in accord with past efforts \citep[e.g.,][]{Mattila05,Leonard07,Boty18}.

For setting limits on narrow H$\alpha$ emission, we mimic the methodology of \citet{Sand18_2017cbv} which we briefly describe here.  We take the flux-calibrated, extinction and redshift-corrected spectra and bin each to their native resolution, between $\approx$1.6-19 \AA.  We then establish a `continuum level' over the broad [\ion{Co}{3}] emission feature and surrounding regions by smoothing the spectrum on scales larger than the expected narrow H$\alpha$ emission, using a second-order Savitsky-Golay filter with a width between 120-180 \AA,~depending on the spectrum.  We experimented with this scale in order to best recover simulated, faint H$\alpha$ features in our data.  Any H$\alpha$ feature of the scale we are interested in would be apparent in the difference between the smoothed and binned spectrum.

To estimate the maximum H$\alpha$ emission that could go undetected in our data, we assume a H$\alpha$ line with FWHM~=~1000 km~s$^{-1}$ and with a peak flux that is three times the root mean square of our residual spectrum, after taking the difference between the smoothed and binned data.  The resulting flux and luminosity limits for these new measurements are presented in Table~\ref{tab:litsearch2} alongside those of previous work.  In Figure~\ref{fig:limits} we plot the H$\alpha$ luminosity limits measured here as a function of $\Delta$m$_{15}(B)$, along with all of the literature measurements compiled in Table~\ref{tab:litsearch2}. Also plotted is the H$\alpha$ detection of ASASSN-18tb, which is significantly stronger than nearly all the limits reported to date -- if emission like that seen in ASASSN-18tb were common it would have been detected in most searches.  To illustrate this even more clearly, we have implanted a simulated line with the luminosity of H$\alpha$ (along with FWHM=1000 km s$^{-1}$ and a velocity offset of 1000 km s$^{-1}$) seen in ASASSN-18tb into our SN~2007gi data, as can be seen in the right panel of Figure~2 -- clearly any line near this strength would stand out in our data.

There is a narrow H$\alpha$ feature seen in the spectrum of SN2003gs at the exact rest frame position of H$\alpha$, as is pointed out in the zoomed in portion of Figure~\ref{fig:spectra}.  The width of the feature is at the resolution of Keck/ESI; $\approx$1.6 \AA~or $\approx$75 km s$^{-1}$.  Upon request, the 2D spectrum of SN2003gs was provided and inspected (T. Brink, private communication), and there was clear, spatially extended H$\alpha$ emission beyond the trace of the SN.  We therefore conclude that this narrow H$\alpha$ emission is from the host galaxy, and not the SN.


\section{Stripped Mass Limits}

In the previous section, we have measured H$\alpha$ luminosity limits for a sample of nebular SN Ia with $\Delta$m$_{15}(B)~>~$1.3 mag, a region of parameter space relatively unexplored by previous work.  In the single degenerate scenario, it is expected that some hydrogen rich material is stripped from the nondegenerate companion star during the explosion, which should manifest as a narrow H$\alpha$ line at late times -- but our search, like most before it, turned up no definitive detections.  Here we translate these line luminosity limits to limits on the amount of stripped hydrogen that could have gone undetected in our observations.  For this, we use the 3D radiative transport results of \citet{Boty18}, whose work derived from the SN Ia ejecta-companion interaction simulations of \citet{Boehner17}.  A simulated spectrum of a normal SN Ia at +200 days after peak brightness was generated, incorporating stripped material from a solar abundance companion star.  For models including main sequence, subgiant and red giant companion stars between $\sim$0.2--0.4 $M_{\odot}$ of material was stripped, leading to $L_{H\alpha}$$\approx$4.5--15.7$\times$10$^{39}$ ergs s$^{-1}$.  These $H\alpha$ luminosities are $\sim$3 orders of magnitude brighter than our measured luminosity limits, and so we can rule out models like that presented by \citet{Boty18}.

\begin{figure*}
\begin{center}
\includegraphics[width=17cm]{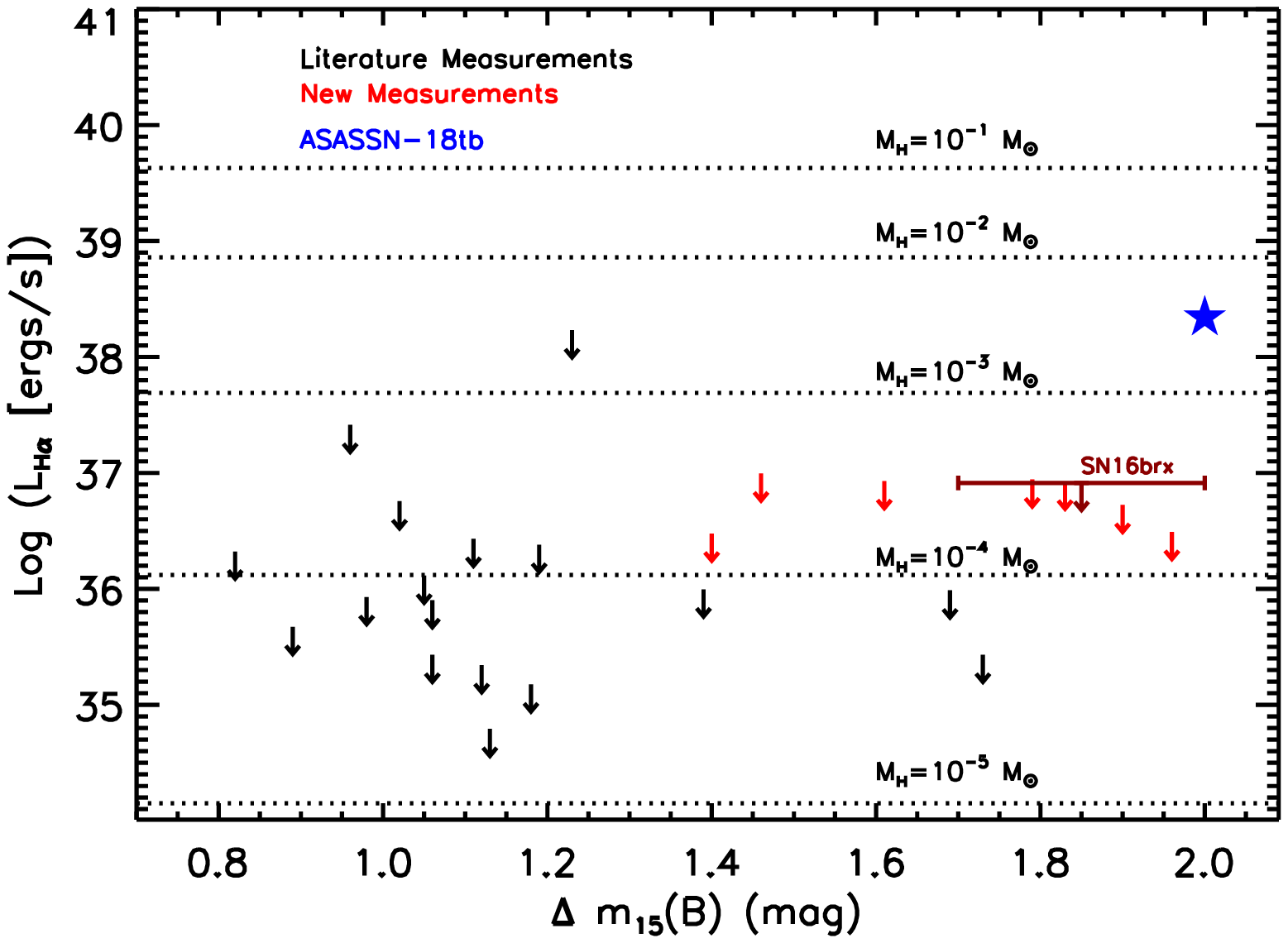}
\caption{Summary of H$\alpha$ luminosity limits from all literature measurements (black upper limits) as well as those presented in this work (red upper limits) at high decline rates.  Since we do not have a $\Delta$m$_{15}(B)$ measurement for SN~2016brx, but it is a confirmed SN91bg-like SN Ia, we have placed a range of 1.7$<$$\Delta$m$_{15}(B)$$<$2 on this object, corresponding to the observed range for this subtype \cite[see e.g. Figure 1 of][]{Taubenberger17}. Also shown is the H$\alpha$ detection in ASASSN-18tb \citep[blue star]{Kollmeier19}, which is significantly stronger than the H$\alpha$ luminosity limits presented here.  Using Eqn 1 from \citet[][with the corrected coefficients from \citealt{Sand18_2017cbv}]{Boty18} we also show the correspondence between stripped hydrogen mass and H$\alpha$ luminosity in their latest set of hydrodynamic simulations.  For all of their single degenerate models, several tenths of a solar mass of hydrogen material was stripped from the companion star.  No attempt was made to correct H$\alpha$ luminosities to the epoch of the \citet{Boty18} models; this order unity effect does not change the conclusions of this plot.   The H$\alpha$ upper limits for all the newly presented fast-declining SN Ia are at least an order of magnitude below the H$\alpha$ detection in ASASSN-18tb.
\label{fig:limits}}
\end{center}
\end{figure*}

If the stripped hydrogen mass is intrinsically lower than that predicted by the \citet{Boty18} models, whether it be due to a weaker explosion, larger companion separation or otherwise, a lower H$\alpha$ luminosity is possible.  To set mass limits on our own data, we use the quadratic formula presented by \citet[see also the typographical correction to this formula in \citealt{Sand18_2017cbv}]{Boty18}, which was derived after manually changing the hydrogen density in their fiducial model.  Using this, we find hydrogen mass limits of $\sim$1--3$\times$10$^{-4}$ M$_{\odot}$.  To put these values in context with other H$\alpha$ limits, and the recent detection in ASASSN-18tb, we compile all the measurements in Figure~\ref{fig:limits}.  Lines of constant hydrogen mass are plotted using the formula provided by \citet{Boty18}.
Note that even though the model nebular spectrum of \citet{Boty18} was generated at +200 days after peak, we make no further correction to our derived hydrogen mass limits even though our spectra (and those in the literature) were taken at other times -- any correction would be of order unity, while the hydrogen limits we present are orders of magnitude below the model expectations.  That said, further modeling efforts are necessary, especially for the weaker explosion energies of fast declining SN Ia, to put definitive limits on the amount of ablated hydrogen mass.  What is robust, however, is the fact that we do not find any other fast declining SN Ia with H$\alpha$ luminosities within $\sim$1--2 orders of magnitude of that found in SN~2018fhw.

\section{Summary and Conclusions}\label{sec:conc}

Nebular H$\alpha$ emission is an important signpost for the single degenerate scenario \citep[e.g.,][]{Marietta00,Mattila05,Boty18}, although it has rarely been detected \citep[but see the sample of circumstellar-interacting SN Ia, whose H$\alpha$ emission appears prior to the nebular phase and may have a different origin; e.g.,][]{Silverman13,Dilday12,Graham19}.  Motivated by recent observations of the fast declining ASASSN-18tb ($\Delta$m$_{15}(B)$~=~2.0 mag), and the conspicuous detection of H$\alpha$ in its nebular spectrum, we have collected all of the H$\alpha$ nebular spectroscopy limits for SN Ia reported in the literature.  Plotting this sample of SN Ia as a function of their decline rate, $\Delta$m$_{15}(B)$, clearly shows a deficit of measurements for faster declining events, $\Delta$m$_{15}(B)~\gtrsim~$1.3 mag.  The origin of this apparent bias is unknown, although it may simply be that faster declining events are intrinsically fainter and are thus harder to observe at late times.  Whatever the origin, it is important to measure nebular H$\alpha$ limits for SN Ia across their decline rate distribution, and amongst their rarer subtypes.  The current work is one step in that direction.

We have added eight SN Ia to the sample with H$\alpha$ limits in their nebular spectra, all with decline rates of $\Delta$m$_{15}(B)~>~$1.3 mag.  For two of these events we present new observations, while the other six have been published elsewhere for other purposes.  No new H$\alpha$ detections were made, with limits  comparable to others in the literature, $\sim$3 orders of magnitude below expectations for the single degenerate model presented by \citet{Boty18}, and $\sim$1--2 orders of magnitude below the recent detection in SN~2018fhw.  Further modeling of SN Ia with varying companion types, separations and explosion energies are necessary before the observations presented here can firmly rule out the single degenerate scenario.

It is interesting to note that all of the fast declining SN Ia with $\Delta$m$_{15}(B)~>~$1.4 mag that we investigated exploded in early-type host galaxies consistent with previous studies that find both transitional and 91bg-like SN Ia favor lenticulars/ellipticals \citep[e.g.,][among others]{Hamuy00,Howell01,Sullivan06,Neill09,Pan14}.  Binary population synthesis studies have found it extremely difficult for single degenerate SN Ia to significantly contribute to the overall SN Ia rate at long delay times \citep[][]{Ruiter09,Mennekens10,Claeys14}.
Indeed, from our current study we have shown that among the fastest declining and most subluminous SN Ia, no more than $\lesssim$15\% exhibit nebular H$\alpha$ emission, with upper limits on H$\alpha$ luminosity which are $\sim$1--2 orders of magnitude below the actual detection in ASASSN-18tb.  

While other studies will explore ASASSN-18tb in more detail, there are at least two things that make it unique and point to future areas of progress.  First,  given its $\Delta$m$_{15}(B)~=~$2.0 mag and its peak absolute luminosity of M$_{B,max}$ = $-$17.7 mag, ASASSN-18tb sits in between the  transitional and 91bg-like SN Ia loci in that parameter space \citep[see Figure 1 of][]{Taubenberger17}; as discussed in Section~\ref{sec:search} there is diversity among the subluminous SNe~Ia and they may originate via multiple channels. Also, ASASSN-18tb exploded in a satellite dwarf elliptical galaxy that is most likely metal poor (ASASSN-18tb's host, LEDA~330802, is projected $\approx$\,4.0 arcminutes, or $\approx$\,80 kpc, from the lenticular galaxy LEDA 14822, which both have nearly the same redshift of $z$ = 0.0175 while LEDA~14822 is 4.2 mag brighter in J-band\footnote{http://ned.ipac.caltech.edu/}).  The mass retention efficiency of accreting white dwarfs is higher at lower metallicity \citep{Shen07,Kobyashi15}, possibly making the single degenerate scenario at long delay times more viable in metal-poor environments. Future surveys for late-time H$\alpha$ should not only cover the $\Delta$m$_{15}(B)$ light curve parameter space as investigated in this study, but also as a function of SN Ia spectral subtype, host galaxy age, and host galaxy metallicity.



\begin{deluxetable*}{lccccccccccc}
\tabletypesize{\scriptsize}
\tablecaption{Nebular spectroscopy log.\label{tab:newspec}}
\tablehead{\colhead{SN Name} & \colhead{Observation} & \colhead{Phase}  & \colhead{Telescope} & \colhead{Grating/Grism}  & \colhead{Slit}            &  \colhead{Exposure} & \colhead{Res} & \colhead{Scaled Mag} & \colhead{E(B-V)$_{MW}$} & \colhead{E(B-V)$_{host}$} & \colhead{Distance} 
\\
\colhead{} & \colhead{Date (UT)}   & \colhead{[days]\tablenotemark{*}} &    \colhead{Instrument}        & \colhead{}          & \colhead{Width [\arcsec]}         & \colhead{Time [s]} & \colhead{(\AA)} &\colhead{(mag)} & \colhead{(mag)} & \colhead{(mag)} & \colhead{(Mpc)}}
\startdata
\hline
\multicolumn{11}{c}{Archival Data}\\
\hline
SN 1999by\tablenotemark{a} & 1999 Nov 09 & +183 & Keck/LRIS & 400/8500 & 1.0" & 250 & 6.9 & R=19.70 & 0.016 & 0.0 & 14.1 \\
SN 2003hv\tablenotemark{b} & 2004 July 25 & +320 & VLT/FORS1 & 300V/I & 1.3" &4$\times$1200 & 11.5 & R=21.13 & 0.016 & 0.0 & 18.79\\
SN 2003gs\tablenotemark{c} & 2004 Feb 14 & +200 & Keck/ESI & Echelle & 1.0" & 1800 & 1.6 & R=20.19 & 0.035 & 0.031 & 21.4\\
SN 2004eo\tablenotemark{d} & 2005 May 16 & +228 & VLT/FORS1 & 300V & 1.0" & 2$\times$2280 & 11.5 & R=21.81 & 0.109 & 0.0 & 67\\ 
SN 2007on\tablenotemark{e} & 2008 Aug 27 & +286 & Mag/LDSS3 & VPH-B & 0.75" & 4$\times$1800 & 2.0 & R=21.59 & 0.0 & 0.0 & 17.9 \\
SN 2016brx\tablenotemark{f} & 2016 Oct 20 & +183 & Mag/LDSS3       & VPH-All     & 1.0\arcsec    & 7$\times$1800 & 10.0 & R=22.4 & 0.083 & 0.0 & 58.2  \\ 
\hline
\multicolumn{11}{c}{New Data}\\
\hline
SN 2007gi & 2008 Mar 26 & +225 & GN/GMOS        & R400     & 0.75\arcsec    & 2$\times$650 & 2.2 & R=20.59 & 0.024 & 0.20 & 24.6  \\ 
SN2017fzw & 2018 Feb 06 & +167 & SALT/RSS & PG0300 & 1.5" & 2226 & 19 & r=19.9 & 0.037 &  0.0 & 25.6\\
 & 2018 Apr 13 & +233 & GS/GMOS &R400+B600 & 0.75" &4$\times$300 & 5.1/5.8 & r=21.6  & '' & '' & '' 
\enddata
\tablenotetext{*}{Phase is in units of days with respect to $B$-band maximum.}
\tablenotetext{a}{\cite{Silverman12}}
\tablenotetext{b}{\cite{Leloudas09} is the source of the spectrum, color excess, distance and late time light curve.}
\tablenotetext{c}{\cite{Silverman12} is the source for the spectrum, while \cite{Krisciunas09} is the source for the color excess, distance and late time light curve.}
\tablenotetext{d}{\cite{Pastorello07} is the source of the spectrum, color excess, distance and late time light curve.}
\tablenotetext{e}{\cite{Folatelli13} is the source for the spectrum, while \cite{Gall18} is the source for the color excess and distance. Note that we used SN~2007gi and SN~2003gs to infer the late time magnitude of SN~2007on, see Section~\ref{sec:data} for details.}
\tablenotetext{f}{\cite{Dong18} is the source of the spectrum and the light curve from which we estimated the late-time magnitude of SN~2016brx.}
\end{deluxetable*}

\acknowledgments
We are grateful to T. Brink for providing us with the 2D spectrum of SN~2003gs, and clarifying the exposure time for SN~1999by.

Based on observations obtained at the Gemini Observatory, which is operated by the Association of Universities for Research in Astronomy, Inc., under a cooperative agreement with the NSF on behalf of the Gemini partnership: the National Science Foundation (United States), National Research Council (Canada), CONICYT (Chile), Ministerio de Ciencia, Tecnolog\'{i}a e Innovaci\'{o}n Productiva (Argentina), Minist\'{e}rio da Ci\^{e}ncia, Tecnologia e Inova\c{c}\~{a}o (Brazil), and Korea Astronomy and Space Science Institute (Republic of Korea) (Programs GN-2008A-Q-17,GS-2018A-Q-315).

This research has made use of the NASA/IPAC Extragalactic Database (NED),
which is operated by the Jet Propulsion Laboratory, California Institute of Technology,
under contract with the National Aeronautics and Space Administration.

Research by DJS is supported by NSF grants AST-1821987 and 1821967. Observations using Steward Observatory facilities were obtained
as part of the observing program AZTEC: Arizona Transient
Exploration and Characterization, which receives support from
NSF grant AST-1515559.  Data from the Southern African Large Telescope were obtained with Rutgers University proposal 2017-1-MLT-002 (PI: SWJ) and supported by NSF award AST-1615455.  P. Milne and M. Moe acknowledge financial support from NASA grant ADAP-80NSSC19K0578.

MLG acknowledges support from the DIRAC Institute in the Department of Astronomy at the University of Washington. The DIRAC Institute is supported through generous gifts from the Charles and Lisa Simonyi Fund for Arts and Sciences, and the Washington Research Foundation.

\vspace{5mm}
\facilities{Gemini North (GMOS), Southern African Large Telescope (RSS) }


\software{
astropy \citep{2013A&A...558A..33A,astropy},   
The IDL Astronomy User's Library \citep{IDLforever}
          }

\bibliographystyle{aasjournal}
\bibliography{biblio}

\end{document}